\documentclass[journal]{IEEEtran}
\usepackage{cite}
\usepackage{amsmath,amssymb,amsfonts}
\usepackage{array, multirow, booktabs}
\usepackage{graphicx}
\usepackage{textcomp}
\usepackage{xcolor}
\usepackage{url}
\usepackage{makecell}
\usepackage{algorithm}
\usepackage{algpseudocode}
\usepackage{subfigure}
\usepackage{placeins} 
\usepackage[left=1.62cm,right=1.62cm,top=1.9cm]{geometry}
\usepackage{etoolbox}

\makeatletter
\g@addto@macro\normalsize{%
  \setlength{\dbltextfloatsep}{6pt plus 1pt minus 1pt}
  \setlength{\dblfloatsep}{2pt plus 1pt minus 1pt}
  \setlength{\@dblfptop}{0pt}
  \setlength{\@dblfpbot}{0pt}
}
\makeatother
\begin{document}

\title{A Novel and Practical Universal Adversarial Perturbations against Deep Reinforcement Learning based Intrusion Detection Systems
}

\author{Hongsen Zhang$^{1}$, Lu Zhang$^{2}$, Gregory Epiphaniou$^{1}$, Carsten Maple$^{1}$,\\$^{1}$WMG, University of Warwick, Coventry, UK\\$^{2}$School of Mathematics and Computer Science, Swansea university, Swansea, UK}

\maketitle

\begin{abstract}
Intrusion Detection Systems (IDS) play a vital role in defending modern cyber physical systems against increasingly sophisticated cyber threats. Deep Reinforcement Learning-based IDS, have shown promise due to their adaptive and generalization capabilities. However, recent studies reveal their vulnerability to adversarial attacks, including Universal Adversarial Perturbations (UAPs), which can deceive models with a single, input-agnostic perturbation. In this work, we propose a novel UAP attack against Deep Reinforcement Learning (DRL)-based IDS under the domain-specific constraints derived from network data rules and feature relationships. To the best of our knowledge, there is no existing study that has explored UAP generation for the DRL-based IDS. In addition, this is the first work that focuses on developing a UAP against a DRL-based IDS under realistic domain constraints based on not only the basic domain rules but also mathematical relations between the features. Furthermore, we enhance the evasion performance of the proposed UAP, by introducing a customized loss function based on the Pearson Correlation Coefficient, and we denote it as Customized UAP. To the best of our knowledge, this is also the first work using the PCC value in the UAP generation, even in the broader context. Four additional established UAP baselines are implemented for a comprehensive comparison. Experimental results demonstrate that our proposed Customized UAP outperforms two input-dependent attacks including Fast Gradient Sign Method (FGSM), Basic Iterative Method (BIM), and four UAP baselines, highlighting its effectiveness for real-world adversarial scenarios.
\end{abstract}


\section{Introduction}
\IEEEPARstart{C}{yber} attacks have long posed a persistent and evolving challenge to the security and stability of modern cyber physical systems (CPS)~\cite{khraisat2019survey}. As technological advancements accelerate, a wide array of intelligent devices has been increasingly integrated into both industrial operations and daily life, substantially enlarging the attack surface and introducing additional avenues for cyber attacks. The growing prevalence and sophistication of such attacks have imposed substantial risks on critical infrastructures including healthcare systems~\cite{ moukafih2024semi}, and even national security~\cite{alkharman2024cyber}. Given these challenges, Intrusion Detection Systems (IDS) have emerged as a crucial component in detecting and mitigating these threats by monitoring network traffic to identify potential malicious activities. IDS can be categorized into Host-based IDS (HIDS) and Network-based IDS (NIDS), depending on whether they analyze host-level or network-level data. However, NIDS is more widely adopted due to its broader coverage and scalability. 

With their great pattern recognition and generalization capabilities~\cite{jordan2015machine}, Machine Learning (ML) and Deep Learning (DL) are increasingly replacing traditional rule-based systems and are now regarded as the core techniques in enhancing the detection performance of NIDS. In this context, Deep Reinforcement Learning (DRL) has gained increasing attention by integrating the generalization strength of DL with the adaptive decision-making capabilities of Reinforcement Learning (RL)~\cite{sutton1998reinforcement}. This combination enables DRL-based IDS to effectively handle large-scale network data without relying on complex feature engineering, while dynamically learning optimal detection policies by interacting with network environments~\cite{lopez2020application,lopez2017conditional, merzouk2022evading}.

Recent research, however, has revealed that despite the remarkable success and generalization capabilities of DL models, they are vulnerable to adversarial attacks~\cite{goodfellow2014explaining, zhang2021countermeasures, zhang2022hybrid}. Adversarial attacks are carefully crafted inputs, typically indistinguishable from legitimate data, that can deceive ML/DL models into making incorrect predictions. When it comes to the intrusion detection area, by introducing minimal adversarial perturbations such as on packet length and size, attackers manipulate the decision-making process of IDSs, severely compromising their reliability and effectiveness. For example, the authors of~\cite{rigaki2017adversarial} proposed FGSM~\cite{goodfellow2014explaining} and Jacobian-based Saliency Map Attack (JSMA)~\cite{papernot2016limitations} to perturb the network flow data for an IDS, which reduced the intrusion detection performance. The authors of~\cite{wang2018deep} further expanded the adversarial attack method on IDS to DeepFool~\cite{moosavi2016deepfool} and Carlini \& Wagner (C\&W)~\cite{carlini2017towards} attack. In addition, Sethi et al.~\cite{sethi2020context} introduced JSMA attack on a DRL-based IDS to evaluate its adversarial vulnerability. However, a notable limitation of these mentioned adversarial approaches is that the generated adversarial attacks are input-dependent, requiring separate perturbation generation for each input instance, which becomes extremely time-consuming and computationally intensive when facing the high-dimensional or large-scale network data. Considering real-time constraints in the deployment of adversarial attacks in NIDS, the authors of~\cite{sheatsley2022adversarial} proposed a Universal Adversarial Perturbation (UAP) attack. This UAP represents a single, input-agnostic perturbation which can be pre-generated and capable of misleading the model across multiple different inputs simultaneously~\cite{moosavi2017universal}. This can be particularly threatening when deployed against target IDS. Nonetheless, an important drawback lies in the fact that, unlike the data in the Computer Vision (CV) domain, network flow data typically follows domain-specific distribution patterns and holds the relationships among different features, which make the attacks impractical or unrealistic if applying the UAP attacks in the CV area to NIDS directly. 

Hence, to address the above challenges, we propose a practical UAP attack against a DRL-based IDS under well-defined realistic domain constraints that incorporate domain-specific rules derived from network data, as well as the mathematical relationships among the features. To further enhance the evasion performance, we advance the proposed UAP attack by taking into consideration the Pearson Correlation Coefficient (PCC) of the predicted outputs between the perturbation and the generated adversarial UAP attack, and we denote it as Customized UAP. PCC quantifies the degree of linear association between two variables, making it a useful measure for assessing the similarity in their patterns or trends. A perturbation with a higher PCC value indicates a stronger alignment in output behavior with that of the generated UAP attack, making it more dominant in shaping the model’s predictions and achieving better attack performance. Additionally, to assess the performance improvements of our Customized UAP attack, we conducted an extensive comparative analysis against four established UAP baselines, as well as the FGSM and BIM attacks. In summary, the contributions of this paper are shown below:

\begin{enumerate}
  \item We propose a practical UAP attack against a DRL-based IDS considering the domain constraints including domain rules of network intrusion detection data and mathematical relationships among their features. To the best of our knowledge, there is no prior study that has explored UAP generation for the DRL-based IDS. In addition, this is also the first work to develop a UAP against DRL-based IDS which formalizes the mathematical relationships among network flow data features in the literature.
  \item We further propose a novel Customized UAP attack by considering PCC between the classification results of the perturbation and the proposed UAP attack. To the best of our knowledge, this is the first work to incorporate PCC into UAP generation, even in a broader context.
  \item Through extensive experimental evaluation, we demonstrate that our proposed Customized UAP attack achieves superior evasion performance as compared to four established UAP baselines, as well as the FGSM and BIM attacks.
\end{enumerate}

The remainder of this paper is structured as follows: Section \ref{Chap:2} provides a comprehensive review of related literature, outlining the recent advances. In Section \ref{Chap:3}, we present a detailed methodology for this work, including the entire process from training the DRL-based IDS to launching adversarial attacks. Section \ref{Chap:4} demonstrates the experimental results with analysis, presenting detailed performance evaluations and insights. Finally, Section \ref{Chap:5} summarizes the primary findings of this work and highlights directions for future research.
\vspace{-1pt}
\section{Literature review} \label{Chap:2}
In this section, we provide a comprehensive literature review of the related works including the taxonomy of IDS, the development of DL in the intrusion detection area, the application of RL and the threat of adversarial attack in the network intrusion detection area. 
\subsection{Network Intrusion Detection System}
An NIDS can be defined as a tool designed to monitor network traffic flow and automatically detect malicious activities within a computer network or system by analyzing traffic patterns~\cite{khraisat2019survey}. Based on their detection approach, NIDS are generally classified into Signature-based Intrusion Detection System (SIDS) and Anomaly-based Intrusion Detection System (AIDS). SIDS could identify and detect known attacks by matching the monitored network flow pattern with some existing attack signatures in the signature database. DL architectures like Deep Neural Network (DNN)\cite{aminanto2016deep}, Convolutional Neural Network (CNN)\cite{vinayakumar2017applying}, and Recurrent Neural Network (RNN)\cite{lopez2017network} are commonly utilized in SIDS to effectively match known attack signatures within high-dimensional network traffic data. In contrast, AIDS learn normal traffic patterns and flag significant deviations as potential attacks, making them more suitable for detecting previously unseen zero-day attacks. Accordingly, techniques related to outlier detection such as Autoencoders~\cite{lopez2017conditional} are preferred in AIDS.

Various datasets have been developed to support IDS training and evaluation, with KDD-Cup99~\cite{kdd_cup_1999_data_130} and its refined version NSL-KDD~\cite{tavallaee2009detailed} being among the most widely used benchmarks. However, they are no longer representative of modern network environments, as they lack contemporary traffic types, updated attack vectors, and realistic traffic diversity. A newer dataset, UNSW-NB15~\cite{moustafa2015unsw} was proposed, which contains contemporary network traffic with a wide range of modern attack types. Similarly, CICIDS2017 and its extended version CICIDS2018~\cite{sharafaldin2018toward} provide labeled traffic mimicking modern network conditions with both benign and diverse contemporary attacks. Moreover, Bot-IoT~\cite{ahmed2018bot} was developed to reflect the attacks targeting Internet of Things (IoT) network environments. Given its comprehensive and up-to-date traffic coverage, we adopt CICIDS2018 to train and evaluate our IDS in this work.

\subsection{Reinforcement Learning}
\label{sec:RL}
RL is a branch of ML where an agent learns to make decisions by interacting with an environment~\cite{sutton1998reinforcement}. Typically, RL is formulated based on a Markov Decision Process (MDP): the agent interacts with the environment in discrete timestep $t$ by first observing the current state $s_t \in S$, then selecting an action $a_t \in A$ according to a decision-making policy $\pi(a_t\mid s_t)$. Once the action is executed, the environment returns a reward $r_t$ or penalty (negative value) and transits the agent to a new state $s_{t+1}$ based on the state transition probability distribution \mbox{$P(s_{t+1} \mid s_t, a_t)$}. By using this trial and error, the agent measures how well it performs certain actions and learns an optimal policy that maximizes the cumulative reward, thus improving decision-making abilities. Early RL algorithms targeted discrete, finite state-action spaces, but real-world problems are often high-dimensional and continuous. Thus, advances in DL have led to the integration of neural networks with RL, forming DRL to address these challenges~\cite{mnih2015human}.

\subsection{Adversarial Attack}
Adversarial attacks were first discovered in the field of CV~\cite{goodfellow2014explaining}, where carefully crafted yet often imperceptible perturbations added to inputs can mislead ML/DL models into incorrect predictions~\cite{zhang2022attention,zhang2025adversarial}. Adversarial attacks can be broadly categorized based on the information that the adversary has for the target model, including white-box, black-box and gray-box attack~\cite{liang2022adversarial}. 

Specifically, in a white-box attack, the adversary has full access to the target model, including its architecture, input data, parameters, and gradients, enabling precise and efficient crafting of adversarial examples via gradient-based methods such as FGSM~\cite{goodfellow2014explaining} and BIM~\cite{kurakin2018adversarial}. FGSM is a one-step, input-dependent attack, and its targeted variant is widely used in adversarial attack research for direct control over misclassification objectives. In this work, we adopt the targeted FGSM, which perturbs the input to minimize the loss with respect to a predefined target label, thereby forcing the model to classify the input as a particular class. FGSM can be mathematically formulated as shown in Equation \ref{equation:5}:
\begin{equation}
\label{equation:5}
\begin{aligned}
x^{\text{adv}} &= x + \delta\\
\text{where\quad} \delta &= - \epsilon \cdot \text{sign} \left( \nabla_x J(\theta, x, y_{\text{target}}) \right) 
\end{aligned}
\end{equation}
in which $x^{\text{adv}}$ means the adversarial example is generated by adding crafted adversarial perturbation $\delta$ to the original input $x$. $\epsilon$ is the attack magnitude controlling the perturbation size under $L_p$-norm, depends on the specific attack scenario. $J(\theta, x, y_{\text{target}})$ is the loss function with respect to a target label $y_{\text{target}}$. $\nabla_x J$ denotes the gradient of the loss with respect to the input $x$. 

On the other hand, BIM~\cite{kurakin2018adversarial} is an iterative extension of FGSM that applies smaller step sizes over multiple iterations, resulting in stronger adversarial examples and improved attack effectiveness than single-step methods. BIM can be mathematically formulated as shown in Equation \ref{equation:6}:
\begin{equation}
\label{equation:6}
\begin{aligned}
x_{n+1}^{\text{adv}} &= x + \delta_{n+1} \\
\text{where\quad} \delta_{n+1} &= \text{Clip}_{\epsilon} \left\{ \delta_n - \alpha \cdot \text{sign} \left( \nabla_x J(\theta, x + \delta_n, y_{\text{target}}) \right) \right\}
\end{aligned}
\end{equation}
\noindent where $\delta_{n+1}$ is the accumulated perturbation at iteration $n+1$, initialized by $\delta_0=0$, and is updated at each step by applying a target FGSM with smaller step $\alpha$. The function $\text{Clip}{\epsilon}(\cdot)$ keeps the accumulated perturbation within the \mbox{$L\infty$-norm} ball of radius $\epsilon$, constraining it to stay in the valid range.

In contrast, a black-box attack assumes the attacker has no knowledge of the model’s architecture or parameters and can only access its output logits or classification results for given inputs, whereas a gray-box attacker has partial knowledge, such as knowing the architecture but not the exact weights, or having access to training data but not the objective function.

From the perspective of perturbation generation, most traditional attacks craft input-specific perturbations, i.e., a unique perturbation for each sample, which can be computationally intensive in large-scale applications. To overcome this, Universal Adversarial Perturbations (UAP)~\cite{moosavi2017universal} aim to generate a single, input-agnostic perturbation capable of fooling the model on most inputs. During an attack, the adversary simply adds the UAP to inputs, avoiding per-sample generation and exposing a severe vulnerability in DNNs.

\subsection{Related works}
\subsubsection*{ML and DL in intrusion detection}

Due to their ability to learn complex patterns from data and outperform traditional heuristic or rule-based methods, ML techniques have attracted considerable attention in intrusion detection~\cite{agrawal2015survey}, enabling automatic detection of subtle anomalies and patterns associated with malicious activities in network traffic~\cite{khraisat2019survey}.
In the early stages, researchers relied on classical ML algorithms to distinguish between benign and malicious activities, such as Logistic Regression~\cite{goeschel2016reducing} and Support Vector Machine~\cite{jha2013intrusion}. Due to the limitations of sensitivity to redundant input features from traditional ML models, these studies often required appropriate feature selection or dimensionality reduction techniques to improve detection performance and model generalization ability~\cite{bajaj2013improving}. 

In recent years, intrusion detection research has increasingly shifted toward DL models, which offer greater flexibility and stronger generalization compared to traditional ML approaches~\cite{aminanto2016deep}. For example, the authors of~\cite{moukafih2020neural} introduced a low-computation DNN-based voting IDS for binary classification, which is suitable for deployment in modern network environments. In addition to DNNs, various DL models have been explored for intrusion detection, with some studies reformulating the task into alternative representations to leverage a broader range of architectures. The work of~\cite{vinayakumar2017applying} proposed a CNN-based IDS to leverage its strength in extracting spatial patterns by transforming network flow data into image-like representations for classification. RNN-based architectures~\cite{lopez2017network} have also been employed in IDS to better capture the temporal evolution features of network traffic data. 
\subsubsection*{DRL and intrusion detection}
Compared to the above ML/DL methods, RL has attracted significant attention for enabling agents to learn optimal behaviors through continuous interaction with dynamic environments, making it particularly suitable for IDS in constantly evolving network environment due to its adaptability and real-time decision-making capabilities. 
The application of RL in IDS can be traced back to 2005, by assigning positive rewards for correct predictions and negative rewards for incorrect ones, updating the weight vector accordingly~\cite{xu2005reinforcement}. Although constrained by limited computational power and data availability, this early work resembled modern RL-based approaches and laid the foundation for subsequent studies. A follow-up study~\cite{servin2005multi} proposed a multi-agent RL-based IDS with a hierarchical architecture in which a central manager coordinated multiple agents and distributed rewards based on classification outcomes. Sethi et al.~\cite{sethi2020context} proposed a context-aware IDS that leverages DRL alongside multiple classifiers. Since the same sample may yield different classification results under varying contexts, the authors employed DRL as a voting mechanism to assist in evaluating the outputs of multiple context-specific classifiers.

A series of studies have integrated DRL into IDS with adversarial environment concepts to enhance training. Based on previous work~\cite{lopez2017conditional}, Caminero et al. proposed a multi-agent DRL framework in which, alongside the classification agent, an environment agent is trained with an opposing reward function so as to select harder-to-classify samples, thereby improving prediction performance and mitigating class imbalance. Under the same framework, they later compared four different DRL agents in another study~\cite{lopez2020application}, each employing a different reward function. According to the experimental results, Double Deep Q-Network demonstrated superior performance.
\subsubsection*{Adversarial example for intrusion detection}
Owing to the flexibility and adaptability offered by RL, DRL has been applied to a wide range of areas in intrusion detection. However, DRL-based IDS also remains under threat due to the vulnerability to adversarial attacks, where imperceptible perturbations, typically constrained under different $L_p$-norms can be added to inputs during inference phase of the IDS to induce misclassification. It is important to emphasize that, unlike in the CV domain where input features (i.e., pixels) are typically independent features, network flow data often exhibit logical dependencies and semantic relationships. As a result, domain-specific constraints must be taken into account to ensure that the generated adversarial examples are valid and meaningful within the network context~\cite{he2023adversarial}.

One of the earliest studies introducing adversarial attacks into the IDS domain~\cite{rigaki2017adversarial} applied FGSM and JSMA in a white-box setting, significantly reducing classifier accuracy. However, the targeted IDS was limited to shallow ML models, and no domain constraints were enforced, allowing unrestricted modification of all features, including binary and categorical features. In addition, the evaluation dataset, NSL-KDD, is outdated. To address the limitations of outdated datasets, Pacheco and Sun~\cite{pacheco2021adversarial} shifted toward using more representative and up-to-date datasets UNSW-NB15 and Bot-IoT. They applied three adversarial attacks against shallow ML models and MLP-based IDSs to compare their resilience to adversarial examples. Nonetheless, beyond the lack of domain-specific constraints, different $L_p$-norm were used for various attack methods, making it difficult to ensure a fair comparison under equal perturbation magnitudes. 

As researchers recognized that unconstrained perturbations can yield impractical adversarial examples in NIDS, more realistic constraints and modification rules have been introduced to ensure their validity and applicability. The work of~\cite{merzouk2022investigating} conducted a comprehensive evaluation of multiple adversarial attacks including FGSM, BIM, DeepFool, C\&W, and JSMA—across three different datasets under various $L_p$-norm constraints. The authors highlighted several key criteria regarding domain-specific constraints for future research: (1) Ensuring modified features remain within valid value ranges; (2) Preserving logical relationships between binary and categorical features; (3) Maintaining semantically valid correlations across certain feature combinations. Swain et al.~\cite{swain2024panda} proposed similar traffic and packet validation constraints to generate practically feasible adversarial examples. The authors applied a mask on specific features, launched a mask-FGSM attack and demonstrated its effectiveness.

The work of~\cite{teuffenbach2020subverting} proposed grouping features by intrinsic properties and ease of modification into four categories: (1) immutable, (2) freely modifiable, (3) modifiable but dependent on other features, and (4) extremely difficult to modify. For the third category, the values must be recalculated according to related features to preserve the underlying semantic and logical relationships between them. This feature-grouping and dependency-aware recalculation approach formalizes domain constraints beyond simple rule-based heuristics, ensuring semantic validity. Similarly,~\cite{debicha2023adv} categorized CICIDS2018 features into modifiable, dependent, and unmodifiable groups, perturbing different combinations of modifiable features to maximize attack effectiveness. The comparison was conducted under the $L_0$-norm, which hindered a fair comparison with other methods at equivalent perturbation levels.

Following the introduction of UAP, researchers have explored their applicability in the IDS domain for real-time adversarial attacks. Sheatsley et al.~\cite{sheatsley2022adversarial} proposed a perturbation histogram method to construct an adversarial sketch, i.e., a UAP based on the key features of adversarial examples generated by JSMA. However, the domain constraints in this study did not account for mathematical dependencies among dimensions. Another work~\cite{10689520} proposed UAP Generator for Industrial IoT (UAPG-IIoT), which generated UAPs based on adversarial examples created by other gradient-based and optimization-based methods. Nonetheless, this approach also did not incorporate feature dependencies.

Regarding adversarial examples for DRL-based IDS, following the prior work~\cite{merzouk2022investigating}, Merzouk et al.~\cite{merzouk2022evading} investigated the vulnerability of a DQN-based IDS to adversarial examples. The authors conducted FGSM and BIM attacks constrained by $L_\infty$-norm and demonstrated that both attacks significantly degraded the IDS's performance, with BIM outperforming FGSM. In a subsequent study~\cite{merzouk2024adversarial}, the authors extended their evaluation by applying FGSM and BIM to five DRL algorithms, revealing robustness differences and the transferability of adversarial examples across DRL-based IDSs. However, the generated perturbations in both two works still involved modifications to all features, without considering the proposed constraints.

To the best of our knowledge, among all existing studies that generate UAP attacks against IDS, none has considered domain constraints that reflect the precise mathematical relationships among input feature dimensions. Furthermore, no existing work has explored UAP generation using a DRL-based IDS agent as target model. Therefore, we are the first to explicitly bridge this research gap, focusing on generating practical and domain-constrained UAP against DRL-based IDS, which provides valuable insight for enhancing the robustness and security of DRL-based intrusion detection systems in real-world scenarios.

\section{Methodology} \label{Chap:3}
In this section, we provide details of our proposed methods. First, we describe the training of the DRL-based IDS, which serves as the target model. Second, we give details of the proposed UAP attack against trained DRL-based IDS considering realistic domain constraints. To the best of our knowledge, this is the first work to conduct a UAP attack targeting DRL-based IDS considering practical domain constraints. Finally, we introduce our Customized UAP, which integrates PCC into the design of the UAP. This further enhances the evasion performance as compared to the proposed UAP attack. To the best of our knowledge, this is the first work to incorporate PCC into the generation process of UAP. To evaluate the effectiveness of our proposed methods, we comprehensively compare them against four established UAP baselines, as well as two input-dependent attacks: FGSM and BIM.
\begingroup
\setlength{\tabcolsep}{12pt}
\setlength{\abovecaptionskip}{-5pt}
\begin{table}[ht]
\caption{Label distribution of training set and testing set}
\label{tab:data_distribution}
\begin{center}
\scalebox{1.1}{
\begin{tabular}{|c|c|c|}
\hline
                & \textbf{Training Set} & \textbf{Testing Set} \\ \hline
\textbf{Benign} & 303,749(50\%)         & 240,702(62\%)        \\ \hline
\textbf{Attack} & 303,749(50\%)         & 147,657(38\%)        \\ \hline
\textbf{Total}  & 607,498               & 388,360              \\ \hline
\end{tabular}}
\end{center}
\end{table}
\endgroup
\vspace{-20pt}
\subsection{DRL agent training} \label{section: data preprocessing and DRL training}
Before giving details of the DRL training, we first present the data preprocessing techniques used in this paper. We choose the CICIDS2018 dataset for its wide range of attack types, comprehensive and realistic representation of modern network traffic, high-quality labeling, and widespread adoption as a benchmark for IDS research~\cite{sharafaldin2018toward}. However, this dataset suffers from class imbalance across attack types and does not provide an official standardized train-test split.
\begin{figure}[t]
\setlength{\abovecaptionskip}{1pt}
\centering
\includegraphics[width=0.39\textwidth, height=0.22\textheight, keepaspectratio=false]{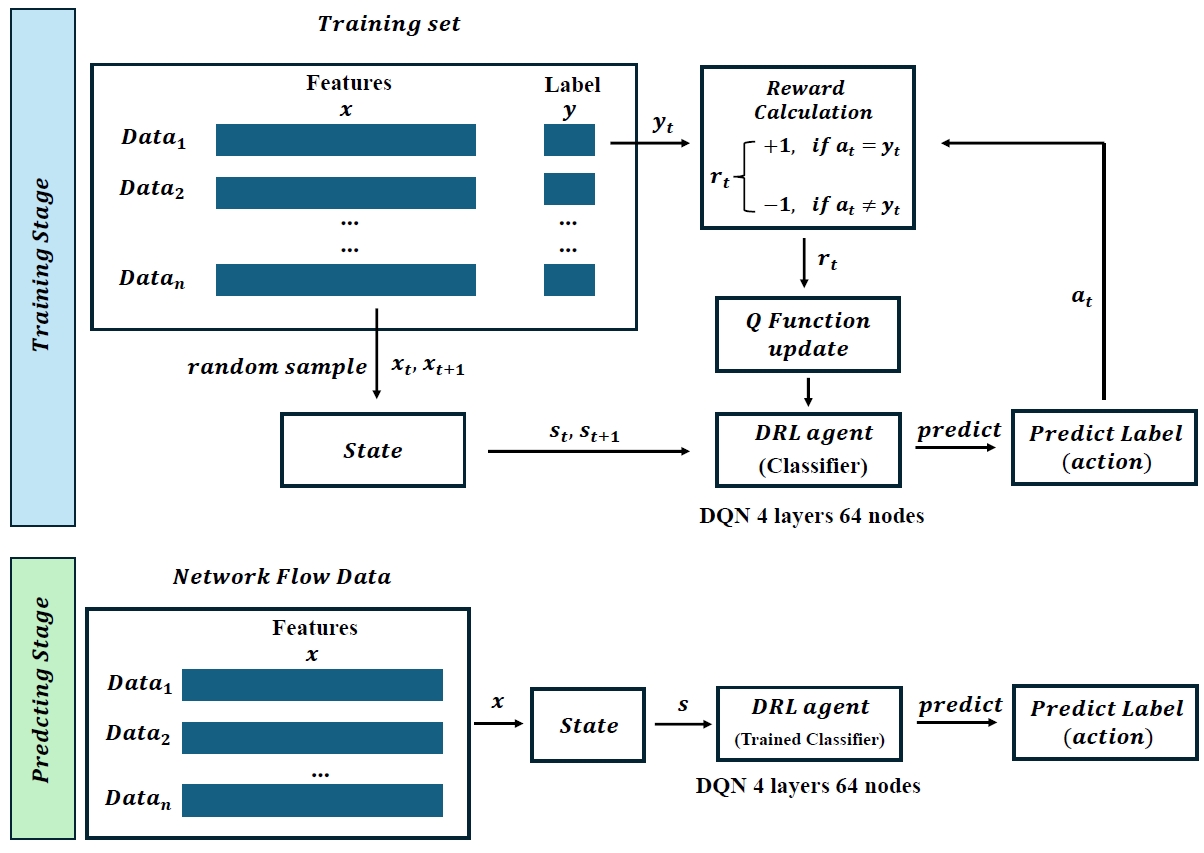}
  \caption{DRL training and predicting stage.}
  \label{fig:DRL training}
\end{figure}
To address these problems, the training and testing sets used in this study are extracted from the raw data files available on the official website. We selected data from two files, "\textit{Friday-02-03-2018\_TrafficForML\_CICFlowMeter.csv}" and "\textit{Friday-16-02-2018\_TrafficForML\_CICFlowMeter.csv}", which include benign samples and three subcategories of attacks. After applying one-hot encoding and Min-Max normalization, the dataset was split into training and testing sets using an 80/20 ratio, while maintaining the same label distribution. To mitigate the class imbalance in the training set, we applied an undersampling strategy, i.e., the number of attack samples was kept unchanged after the split, while benign samples were randomly reduced to match the attack class size. The final training and testing sets have 77 features (including 1 label feature) and the label distributions are shown in Table \ref{tab:data_distribution}.

For DRL agent training, before delving into the DRL process, we first introduce the formalized MDP in IDS problem. In an MDP, the agent interacts with the environment by taking actions in a given state and transits to the next state. Importantly, MDPs rely on the Markov property, assuming that the next state depends solely on the current state and action, enabling the agent to make decisions based only on present information~\cite{sutton1998reinforcement}. The adopted CICIDS2018 dataset is a flow-based dataset, in which each instance represents a distinct network connection, and temporal features have been aggregated into static statistical features. As a result, the data instances are relatively independent of one another, making it appropriate to model the problem as an MDP.
In elucidating the DRL problem in IDS, we define the \textbf{state} as the complete feature set $x$ of each data instance, the DRL agent takes the features (state) as input and outputs the predicted label, which serves as the \textbf{action} in the environment. The \textbf{reward} function $r$ is generated according to the prediction result.  The detailed DRL training stage and predicting stage are shown in \mbox{Figure \ref{fig:DRL training}}.

\begin{figure}[H]
\setlength{\abovecaptionskip}{1pt}
\centering
\includegraphics[width=0.33\textwidth, height=0.14\textheight, keepaspectratio=false]{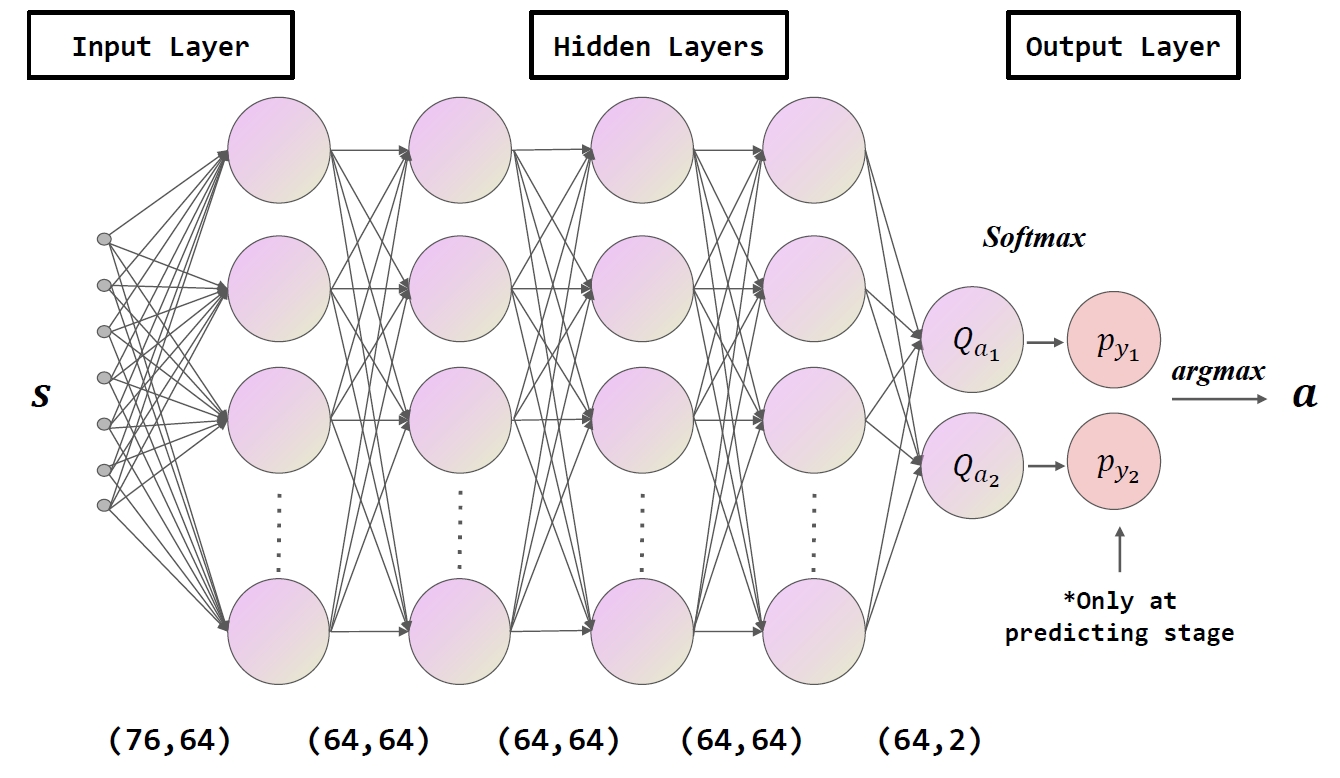}
  \caption{\centering Network architecture of DRL agent.}
  \label{fig:DRL architecture}
\end{figure}

Specifically, during the training stage, the process begins with the training set on the left: Given a timestep $t$, two data samples $x_t$, $x_{t+1}$ will be randomly drawn and assigned as the current state $s_t$ and the next state $s_{t+1}$, respectively. Meanwhile, the label $y_t$ associated with $x_t$ is extracted for later comparison. Then the states are fed into the DRL agent, in this case, we employ a DQN architecture consisting of 4 fully connected layers with 64 nodes each and \textbf{ReLU} as the activation function. The input dimension is 76, corresponding to the feature vector $x$, and the output dimension is 2, representing the number of possible labels $y$, shown in Figure \ref{fig:DRL architecture}.  During the training stage, the activation values of the last hidden layer stand for the $Q$ function of different actions $a$. After obtaining the $Q$-values for each action, the action with the highest $Q$-value is selected as the predicted label. This prediction $a_t$ is subsequently compared with the ground truth label $y_t$ to calculate the reward $r_t$ associated with the current decision. The reward $r_t$ is then used to update the Q-function according to the Bellman equation~\cite{sutton1998reinforcement}, as shown in \mbox{Equation \ref{equation:3}}.
\begin{equation}
\label{equation:3}
Q(s_t, a_t) \leftarrow r_t + \gamma \max_{a'} Q(s_{t+1}, a')
\end{equation}
where $Q(s_{t+1}, a')$ denotes the estimated future return from the next state $s_{t+1}$ under all possible actions $a'$, and $\gamma \in [0, 1]$ is the discount factor controlling the contribution of future rewards, which is set to 0.001 in this work. As the $Q$-function is represented by a neural network, the network parameters $\theta$ are optimized by minimizing the Temporal Difference (TD) error, which serves as the training loss (see Equation \ref{equation:4}). By treating all data in the training dataset as a single episode, each agent is trained over 10 such episodes during the training phase.
\begin{equation}
\label{equation:4}
\mathcal{L} = ( r_t + \gamma \max_{a'} Q(s_{t+1}, a'; \theta) - Q(s_t, a_t; \theta))^2
\end{equation}
 During the predicting phase, as shown in Figure \ref{fig:DRL architecture}, a \textbf{Softmax} layer is added to the trained DRL agent for classification, which is then used as a classifier to predict the labels of incoming network traffic data, as illustrated in the lower part of \mbox{Figure \ref{fig:DRL training}}.
 \vspace{-10pt}
\begingroup
\setlength{\abovecaptionskip}{-1pt}
\begin{table*}[ht]
\caption{Feature categorization, descriptions, and formulations}
\label{tab:feature_groups}
\centering

{ \renewcommand{\arraystretch}{1.3}
\begin{tabular}{|>{\centering\arraybackslash}m{2.3cm} 
                |>{\centering\arraybackslash}m{5cm} 
                |>{\centering\arraybackslash}m{6cm} 
                |>{\centering\arraybackslash}m{3cm}|}
\hline
\makecell{\textbf{Feature Name}} & 
\makecell{\textbf{Description}} & 
\makecell{\textbf{Recalculate Formulation}} & 
\makecell{\textbf{Group}} \\ \hline

Tot Fwd Pkts & Total packets in the forward direction & – & \multirow{5}{*}{Modified Features} \\ \cline{1-3}
Tot Bwd Pkts & Total packets in the backward direction & – & \\ \cline{1-3}
TotLen Fwd Pkts & Total size of packet in forward direction & – & \\ \cline{1-3}
TotLen Bwd Pkts & Total size of packet in backward direction & – & \\ \cline{1-3}
Flow Duration & Duration of the flow in Microsecond & – & \\ \hline

Fwd Pkts/s & Number of forward packets per second & 
${\text{Tot Fwd Pkts} \times 10^6}\;/\;{\text{Flow Duration}}$& 
\\ \cline{1-3}

Bwd Pkts/s & Number of backward packets per second & 
${\text{Tot Bwd Pkts} \times 10^6}\;/\;{\text{Flow Duration}}$ & 
\\ \cline{1-3}

Flow Pkts/s & Number of flow packets per second & 
{$\displaystyle \text{Fwd Pkts/s} + \text{Bwd Pkts/s}$} & 
\\ \cline{1-3}

Flow Byts/s & Number of flow bytes per second & 
\rule{0pt}{17pt}\raisebox{1ex}{$\displaystyle \frac{(\text{TotLen Fwd Pkts} + \text{TotLen Bwd Pkts}) \times 10^6}{\text{Flow Duration}}$} & \raisebox{-1\baselineskip}{Related Features}
\\ \cline{1-3}

Pkt Size Avg & Average size of packet & 
\rule{0pt}{17pt}\raisebox{1ex}{$\displaystyle \frac{\text{TotLen Fwd Pkts} + \text{TotLen Bwd Pkts}}{\text{Tot Fwd Pkts} + \text{Tot Bwd Pkts}}$} & 
\\ \cline{1-3}

Fwd Seg Size Avg & Average size in the forward direction & 
$\text{TotLen Fwd Pkts}\;/\;{\text{Tot Fwd Pkts}}$ & 
\\ \cline{1-3}

Bwd Seg Size Avg & Average size in the backward direction & 
$ {\text{TotLen Bwd Pkts}}\;/\;{\text{Tot Bwd Pkts}}$ & 
\\ \cline{1-3}

Down/Up Ratio & Download and upload ratio & 
{$\text{int} ( {\text{Tot Bwd Pkts}}\;/\;{\text{Tot Fwd Pkts}} )$} & 
\\ \hline

Other Features & Remaining features not in above groups & – & Unmodified Features \\ \hline
\end{tabular}
}
\end{table*}
\endgroup
\vspace{-10pt}
\subsection{The Proposed UAP Attack} \label{section: constraint FGSM/BIM UAP}

In this subsection, we propose a realistic UAP attack considering practical domain constraints. We utilize a feature grouping approach which is inspired by~\cite{teuffenbach2020subverting}. However, the main difference between our approach and the work of~\cite{teuffenbach2020subverting} is that, instead of only considering grouping result, we investigated the underlying mathematical relationships among relevant features to enhance the understanding of their interdependencies and summarized these relationships to derive optimal feature groupings. In addition, we integrated both the domain rules from network flow data and the feature grouping result into the design of our proposed UAP attack. To the best of our knowledge, this is the first work in the literature to develop a UAP against a DRL-based IDS that systematically formalizes the mathematical relationships among network flow data features.

Before delving into details, we will introduce the threat settings we used in this work. First, this work focuses exclusively on evasion attacks at the inference stage, where the adversary perturbs inputs after model training. We consider a white-box threat scenario, where the adversary has full knowledge to the target model’s parameters and gradients. The intrusion detection task is formulated as a binary classification problem, in which malicious traffic is labeled as positive (Label: 1) and benign traffic as negative (Label: 0). For the attacker’s capability and target, the adversary strategically perturbs specific input features under the $L_\infty$-norm constraint, with the objective of causing malicious instances to be misclassified as benign. Importantly, to ensure that the generated adversarial examples are realistic and practically deployable, the attacker’s modifications must obey predefined domain constraints. 

After data preprocessing as discussed in Section \ref{section: data preprocessing and DRL training}, the dataset consists of 76 input features. The input features are categorized into three groups: \textit{Modified features (MF)}, \textit{Related features (RF)}, and \textit{Unmodified features (UF)}. MF means that they can be perturbed arbitrarily by the adversary. However, the perturbations applied must remain within the valid value range of each feature, as the rules introduced in~\cite{merzouk2022investigating}. Specifically, given an input $x$ belongs to the dataset $X$, the constraints for MF are shown in \mbox{Equation \ref{equation:1}}:
\begin{equation}
\label{equation:1}
\forall x \in \mathcal{X},\quad \forall i \in MF,\quad (x + \delta)_i \in \mathcal{R}_i(\mathcal{X})
\end{equation}
where $x_i$ stands for the $i$-th feature of $x$ belong to feature group $MF$, $\delta$ denotes the perturbation for $x$ and $x+\delta$ stands for the adversarial example. This equation depicts that although the values of MF features are perturbed, each modified feature value of the adversarial example is still required to lie within the valid range corresponding to the dataset $\mathcal{X}$. For $RF$ feature group, the domain constraints are illustrated in Equation \ref{equation:2}:
\begin{equation}
\label{equation:2}
\forall x \in \mathcal{X},\quad \forall j \in RF,\quad (x + \delta)_j = \text{Recalculate} (x + \delta)_i\ 
\end{equation}
where the features that belong to $RF$ exhibit explicit mathematical relationships with the features in $MF$. This implies that whenever the adversary modifies any feature in $MF$ (denoted as the $i$-th feature), the corresponding features in $RF$  (denoted as the $j$-th feature) must be recalculated accordingly instead of free-perturbation. The specific form of the \textit{Recalculate} function varies depending on the specific mathematical relationships among the features. 

The rest features are categorized into the $UF$, the values of these features remain fixed throughout the attack process. Since no explicit feature formulation is provided in existing CICIDS2018 resources, we investigate, validate and summarize these relationships among the features. We provide the details of feature grouping and recalculation formulation as shown in Table \ref{tab:feature_groups}. Specifically, based on the feature grouping results, the five features: \textit{Tot Fwd Pkts}, \textit{Tot Bwd Pkts}, \textit{TotLen Fwd Pkts}, \textit{ TotLen Bwd Pkts}, and \textit{Flow Duration} are categorized into the modified feature group. These features span three core aspects of network flow behavior: packet count (\textit{Tot Fwd Pkts}, \textit{Tot Bwd Pkts}), packet size (\textit{TotLen Fwd Pkts}, \textit{TotLen Bwd Pkts}), and flow duration (\textit{Flow Duration}). In real-world scenarios, adversaries can modify all these dimensions through specific actions, such as fragmenting or padding packets, injecting additional traffic, or adjusting the timing of transmissions~\cite{debicha2023adv}. Therefore, ensuring both the individual feature modification and the preservation of their dependencies is practically feasible under this domain constraint.
To evaluate the performance of the proposed UAP attacks, two well-established adversarial attack methods in targeted attack setting were adopted as baselines, namely, FGSM and BIM. In our case, we set \textit{Benign} class as the target label in Equation \ref{equation:5} and \ref{equation:6} since the adversary's target is to mislead the classifier into misclassifying all the malicious instances as benign. The cross entropy loss is used for $J(\theta, x, y_{\text{target}})$ in this work, which is consistent with the proposed UAP attack for a fair comparison. Now we introduce the proposed UAP attack considering realistic domain constraints. The details of the UAP algorithm are given in Algorithm \ref{alg:uap}.

In the initialization stage (lines 1–4), $Seedset$ is randomly sampled from the training set $Tr$, where $size$ defines the proportion of samples randomly selected from the $Tr$ and is set to 0.001 in our experiments. The output $uap$ is initialized as a zero vector with the dimensionality of input features in line~2. During the UAP generation phase, the seedset is shuffled at the 

\begin{algorithm}[H]
\caption{UAP Generation for Classifier $C$}
\label{alg:uap}
\textbf{Input:} Trainset $Tr$, Classifier $C$, Modified feature group mask $mask$, seedset size $size$, expected fooling rate $delta$, max iteration number $max\_iter$, epsilon $eps$ \\
\textbf{Output:} Universal adversarial perturbation $uap$
\vspace{1ex}
\begin{algorithmic}[1]
\State $Seedset \gets$ randomly select $size \times \text{len}(Tr)$ from $Tr$
\State $uap \gets 0$ for all features
\State $fr \gets 0$
\State $iter\_num \gets 0$
\While{$fr < 1 - delta$ and $iter\_num < max\_iter$}
    \State Shuffle $Seedset$
    \For{$data$ in $Seedset$}
        \State $L_1,\;L_2 \gets C(data), \;C(data + uap)$
        \If{$L_1 = L_2$}
            \State $ae \gets$ target FGSM($C$, $data$, $eps$, $L_{\infty}$, $mask$)
            \State $perturbation \gets ae - data$
            \State $uap \gets uap + perturbation$
            \State $uap \gets \text{Project}_{L_{\infty}}(uap)$
        \EndIf
    \EndFor
    \State $iter\_num \gets iter\_num + 1$
    \State $fr \gets \displaystyle  (C(Tr + uap) \neq C(Tr)) \; /  \;\text{len}(Tr) $
\EndWhile
\State \textbf{return} $uap$
\end{algorithmic}
\end{algorithm}
\noindent beginning of each iteration in line 6. The classifier $C$ refers to the pre-trained DRL-based agent. For each data instance in the seedset (line 7), the original prediction $C(data)$ and the perturbed prediction $C(data + uap)$ are computed at line 8. If the two labels are identical, it indicates that the current $uap$ has not successfully altered the model’s prediction. A binary mask $mask$ specifies feature constraints for perturbation generation, with 1 indicating modifiable features and 0 indicating fixed ones. In such cases, a perturbation is generated using the targeted FGSM method with the predefined $mask$, \textit{Benign} label as $y_{target}$, is then added to the current $uap$. After each update (lines 10–12), the accumulated $uap$ is projected back onto the $L_\infty$ norm ball using a projection function to enforce the perturbation constraint in line 13. Then the corresponding iteration number $iter\_num$ and fooling rate $fr$ of this iteration are updated (lines 16–17). The procedure terminates once it achieves the desired fooling rate denoted by $1 - delta$ or reaches the maximum number of iterations $max\_iter$, at which point the final $uap$ is returned.
\subsection{The Proposed Customized UAP}\label{sec:custom_uap}
In this subsection, we present our proposed Customized UAP attack, which incorporates PCC for guiding perturbation generation of the proposed UAP in Section \ref{section: constraint FGSM/BIM UAP}. PCC is a statistical metric that measures the linear correlation between two variables and serves as an effective indicator of similarity in their behavior. Under the guidance of the PCC objective, the generated UAP is optimized to produce greater similarity in prediction outcomes to those of the final adversarial examples, thereby strengthening the perturbation's dominance in the final result and enhancing attack effectiveness. To the best of our knowledge, this is the first attempt to utilize PCC in the context of UAP generation for IDS, even in the broader adversarial learning context. For a more comprehensive evaluation, we also implement several established baseline UAP attacks for comparison. 


Before delivering the details of Customized UAP, we give the detailed definition of PCC. It quantifies the linear relationship between two groups of variables $X$ and $Y$ with $n$ dimensions each (in Equation \ref{equation:7}), ranging from –1 (perfect negative correlation) to +1 (perfect positive correlation). It was first introduced in the context of adversarial attacks in the work of~\cite{zhang2020understanding} in CV area, and has since been adopted as a core measurement~\cite{ weng2023exploring}.
\begin{equation}
\text{PCC}(X, Y) = \frac{\sum_{i=1}^{n} (X_i - \bar{X})(Y_i - \bar{Y})}{\sqrt{\sum_{i=1}^{n} (X_i - \bar{X})^2} \sqrt{\sum_{i=1}^{n} (Y_i - \bar{Y})^2}}
\label{equation:7}
\end{equation}

The authors of~\cite{zhang2020understanding} proposed that both the clean input data $x$ and the perturbation $\delta$ contribute to the final classification outcome of an adversarial example $x + \delta$. If the perturbation is further treated as an independent sample, we can compute the value of the classifier outputs $\text{activation}(x)$, $\text{activation}(\delta)$, and $ \text{activation}(\delta + x)$, respectively. Note that these outputs represent the activation values of the last hidden layer (i.e., logits) rather than the final predicted labels. According to Equation \ref{equation:8}, the authors calculated the values of $\text{PCC}_x$ and $\text{PCC}_{pertu}$, which show the contribution degrees to the prediction result. A higher PCC value indicates more contribution to the final prediction of the adversarial example.
\begin{equation}
\label{equation:8}
\begin{aligned}
\text{PCC}_{x} = \text{PCC}(\text{activation}(x), \text{activation}(\delta + x))\\
\text{PCC}_{pertu} = \text{PCC}(\text{activation}(\delta), \text{activation}(\delta + x))
\end{aligned}
\end{equation}
\begin{figure}[H]
\centering
\setlength{\abovecaptionskip}{1pt}
\includegraphics[width=0.31\textwidth, height=0.18\textheight, keepaspectratio=false]{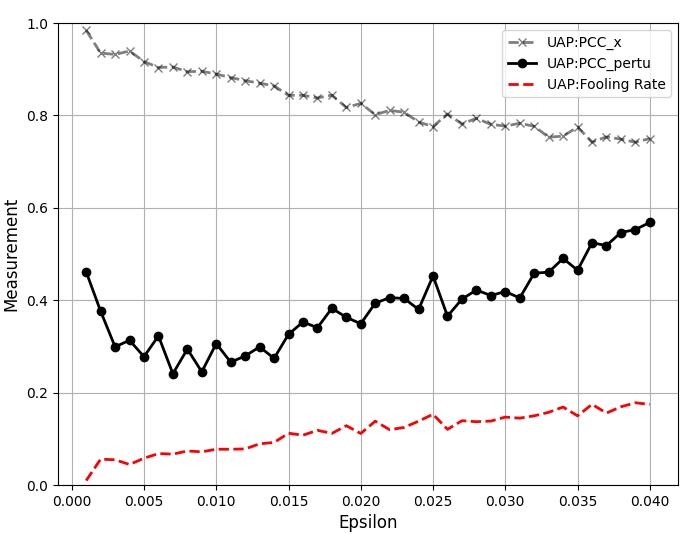}
  \caption{PCC Values and Fooling Rate of Proposed UAP Attack}
  \label{fig:UAP PCC compare}
\end{figure}
\vspace{-5pt}
Here we provide the rationale why we integrate PCC into UAP design. Following Equation \ref{equation:8} above, we further investigate the relation among two PCC values and the attack performance in UAP generation in NIDS. Figure \ref{fig:UAP PCC compare} illustrates the behavior of our proposed UAP attack in Section \ref{section: constraint FGSM/BIM UAP} under the $L_\infty$-norm constraint across a range of epsilon values from 0 to 0.04, averaged over 80 runs. Specifically, it presents the fooling rate and the averaged PCC values over all test instances: $\text{PCC}_x$, $\text{PCC}_{pertu}$. Since $\text{PCC}_{pertu}$, $\text{PCC}_x$ stand for the contribution level to the final prediction from the side of perturbation and clean data, respectively, a higher PCC value on either side indicates a greater influence on the model’s decision, thereby reflecting which component is more dominant in the overall prediction outcome. It can be observed that, as epsilon increases, the fooling rate improves consistently, indicating a better attack effectiveness. We can also find that there exists a positive correlation between the $\text{PCC}_{pertu}$ value of the perturbation and the fooling rate of the UAP. Meanwhile, the two PCC-based metrics exhibit opposite trends: as the fooling rate increases, the $\text{PCC}_{\text{pertu}}$ rises, while $\text{PCC}_x$ decreases correspondingly. This shows that the more dominant the perturbation side to the prediction outcome is, the more effective the UAP attack tends to be. Inspired by this observation, we introduce $\text{PCC}_x$ and $\text{PCC}_{pertu}$ as extra metrics to further assess the intrinsic relationship between the input data $x$ and the adversarial perturbation $\delta$ across the corresponding adversarial examples in our experiment. Drawing upon this understanding, a natural question arises: if a more dominant perturbation leads to better attack performance, can $\text{PCC}_{\text{pertu}}$ itself be used directly as a loss function to generate stronger UAP? 

Now we give the detailed UAP design with $\text{PCC}_{\text{pertu}}$ value. To explore the idea from the question, based on the generation process shown in Algorithm \ref{alg:uap}, we introduce a novel objective function $J$ in line 11, i.e., the $\text{PCC}_{\text{pertu}}$ value, shown in Equation \ref{equation:9}. This encourages the generated perturbation to maximize its contribution to the model’s prediction result, as measured by an increased $\text{PCC}_{\text{pertu}}$ value. 
\begin{equation}
\label{equation:9}
\begin{aligned}
x^{\text{adv}} &= x + \epsilon \cdot \text{sign} \left( \nabla_x J(\theta, x) \right)\\
\text{where\quad} J&= \text{PCC}(\text{activation}(\delta), \text{activation}(\delta + x))
\end{aligned}
\end{equation}

In addition, to make a comprehensive comparison, we also explore the attack performance of established UAP attacks with different loss functions as baselines. The detailed descriptions of each baseline are provided in Table \ref{tab: Milti Loss}, including the loss functions with their names shown in the figures in the next chapter. Notably, since the original works did not assign explicit names to the proposed loss functions, all UAP baselines in the table are named according to their design principles for clarity and consistency in comparison. Additionally, $\text{activation}_{l_i}$ in the table refers to the activation vector at the $i$-th network layer. 

\begin{table*}[t]
\setlength{\dbltextfloatsep}{-100pt}
\setlength{\abovecaptionskip}{-1pt}
\caption{Different Custom Loss Function for Comparison}
\label{tab: Milti Loss}
\centering

{ \renewcommand{\arraystretch}{1.2}
\begin{tabular}{|>{\centering\arraybackslash}m{5cm} 
                |>{\centering\arraybackslash}m{9cm}|}
\hline
\makecell{\textbf{Baseline UAP}} & 
\makecell{\textbf{Loss Function to Maximize}} \\ \hline

PD\_mean\_UAP\cite{mopuri2017fast} & 
\rule{0pt}{19pt}\raisebox{1.1ex}{
\makecell{
$\log\left( \prod_{i=1}^{k} \text{mean}(l_i + \text{eps}) \right)$ \\
$\text{where} \; l_i = \text{activation}(x + \delta),\; \text{eps} = 1 \times 10^{-8}$
}} \\ \hline

PD\_L2\_UAP\cite{mopuri2018generalizable} & 
\rule{0pt}{19pt}\raisebox{1.1ex}{
\makecell{
$\log\left( \prod_{i=1}^{k} \left\| l_i \right\|_2 + \text{eps} \right)$ \\
$\text{where} \; l_i = \text{activation}(x + \delta),\; \text{eps} = 1 \times 10^{-8}$
}} \\ \hline

COSSIM\_L3\_UAP\cite{ye2023fg} &
\rule{0pt}{17pt}\raisebox{1ex}{
\makecell{
$- \text{cossim}\left( \text{activation}_{l_i}(x + \delta),\ \text{activation}_{l_i}(x) \right)$\\
$\quad \text{where} \quad l_i = l_3$
}} \\ \hline

COSSIM\_L4\_UAP\cite{ye2023fg} & 
\rule{0pt}{17pt}\raisebox{1ex}{
\makecell{
$- \text{cossim}\left( \text{activation}_{l_i}(x + \delta),\ \text{activation}_{l_i}(x) \right)$\\
$\quad \text{where} \quad l_i = l_4$
}} \\ \hline

PCC\_pertu\_UAP (\textbf{\textit{our Customized UAP}}) & 
\rule{0pt}{17pt}\raisebox{1ex}{
\makecell{
$\text{PCCsim}\left( \text{activation}_{l_i}(x + \delta),\ \text{activation}_{l_i}(\delta) \right)$\\
$\quad \text{where} \quad l_i = l_4$
}} \\ \hline
\end{tabular}
}
\end{table*}

Among these loss functions, the authors of~\cite{mopuri2017fast} and~\cite{mopuri2018generalizable} propose that, within a classifier, each neuron at every layer is responsible for detecting specific semantic attributes of the input data sample. When a neuron successfully detects its corresponding attribute, it becomes more "active", reflected by a higher output value after activation. Building on this insight, the authors argue that the classifier would be unable to distinguish between clean and adversarial samples if a UAP can misfire all neurons across layers, i.e., to induce abnormally high activations regardless of the true data semantics. To this end, they introduce two distinct loss functions to amplify the activations of all neurons, denoted as \textit{PD\_mean\_UAP} and \textit{PD\_L2\_UAP}, respectively. 

\textit{COSSIM\_L3\_UAP} and \textit{COSSIM\_L4\_UAP}, proposed by Ye et al.~\cite{ye2023fg}, are designed to reduce the similarity of activation values between clean inputs and their corresponding adversarial examples. This approach is motivated by the observation: For adversarial examples, by intentionally misleading the classifier into producing incorrect predictions, not only affects the final output layer (logits) but also alters the internal activation vector, particularly the last several layers of the network. Specifically, these two loss functions apply this strategy at different network layers and use cosine similarity as their measurement. Finally, our proposed Customized UAP is presented as \textit{PCC\_pertu\_UAP}. The experimental results with a comprehensive analysis will be provided in Chapter \ref{Chap:4}.

\section{Experimental Results}\label{Chap:4}

In this section, we first present the training result of DRL-based IDS and the attack performance of FGSM and BIM. Then, we report the experimental results of the proposed UAP attack and Customized UAP. Furthermore, to comprehensively assess the enhanced performance of our Customized UAP, we conducted a comparative analysis against four established UAP baselines as well as two data-dependent methods, i.e., FGSM and BIM.

Before showing the results, we first introduce the metrics we use for comparison. Given that the objective of the attacker is to induce the misclassification of \textit{Attack} (positive) samples as \textit{Benign} (negative). To evaluate the performance of the proposed adversarial attack method, we adopt \textbf{Accuracy} and \textbf{False Negative Rate (FNR)} as the primary evaluation metrics throughout all experiments. Accuracy and FNR are mathematically defined in Equations \ref{equation:10}:
\begin{equation}
\text{Accuracy} = \frac{TP + TN}{total\: samples}\text{,} \quad
\text{FNR} = \frac{FN}{TP + FN}
\label{equation:10}
\end{equation}
Accuracy measures the proportion of correctly classified samples, while FNR indicates the likelihood of misclassifying malicious (positive) samples as benign (negative), reflecting attack success rate. To show the respective contributions of clean data and perturbations to final classification output, $\text{PCC}_x$ and $\text{PCC}_{pertu}$ are also used, as described in \mbox{Equation \ref{equation:8}}.
\subsection{DRL Training and Adversarial Setup}
To facilitate the transparency of the experimental setup, detailed parameter settings for DRL agent training and adversarial attacks (FGSM and BIM) are introduced. All experiments were conducted in Python, with DNNs implemented in PyTorch framework~\cite{paszke2019pytorch}. The training dataset was transformed into an MDP (Chapter~\ref{Chap:3}) and modeled as a custom gymnasium environment~\cite{towers2024gymnasium}, then integrated with Stable-Baselines3~\cite{stable-baselines3} to train the DQN agent shown in Figure~\ref{fig:DRL architecture}. 
\begin{figure}[H]
\setlength{\abovecaptionskip}{1pt}
\centering
\includegraphics[width=0.33\textwidth, height=0.17\textheight, keepaspectratio=false]{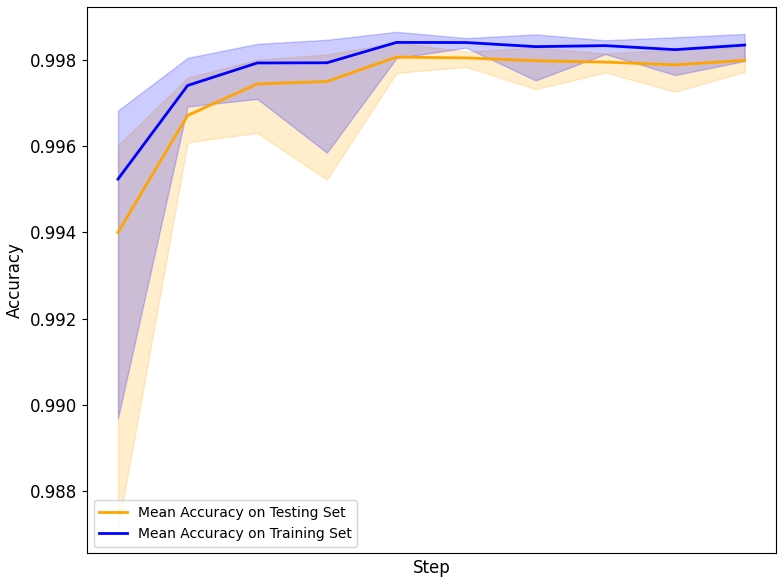}
  \caption{Mean training and testing accuracy across 10 DRL agent training runs}
  \label{fig:DRL performance}
\end{figure}
\begin{figure*}[t]
\setlength{\abovecaptionskip}{1pt}
    \centering
    \subfigure[FNR for FGSM/BIM attack with and without domain constraint, the attack labels include ‘NC’ means no constraint]{
        \includegraphics[width=0.31\textwidth, height=0.18\textheight, keepaspectratio=false]{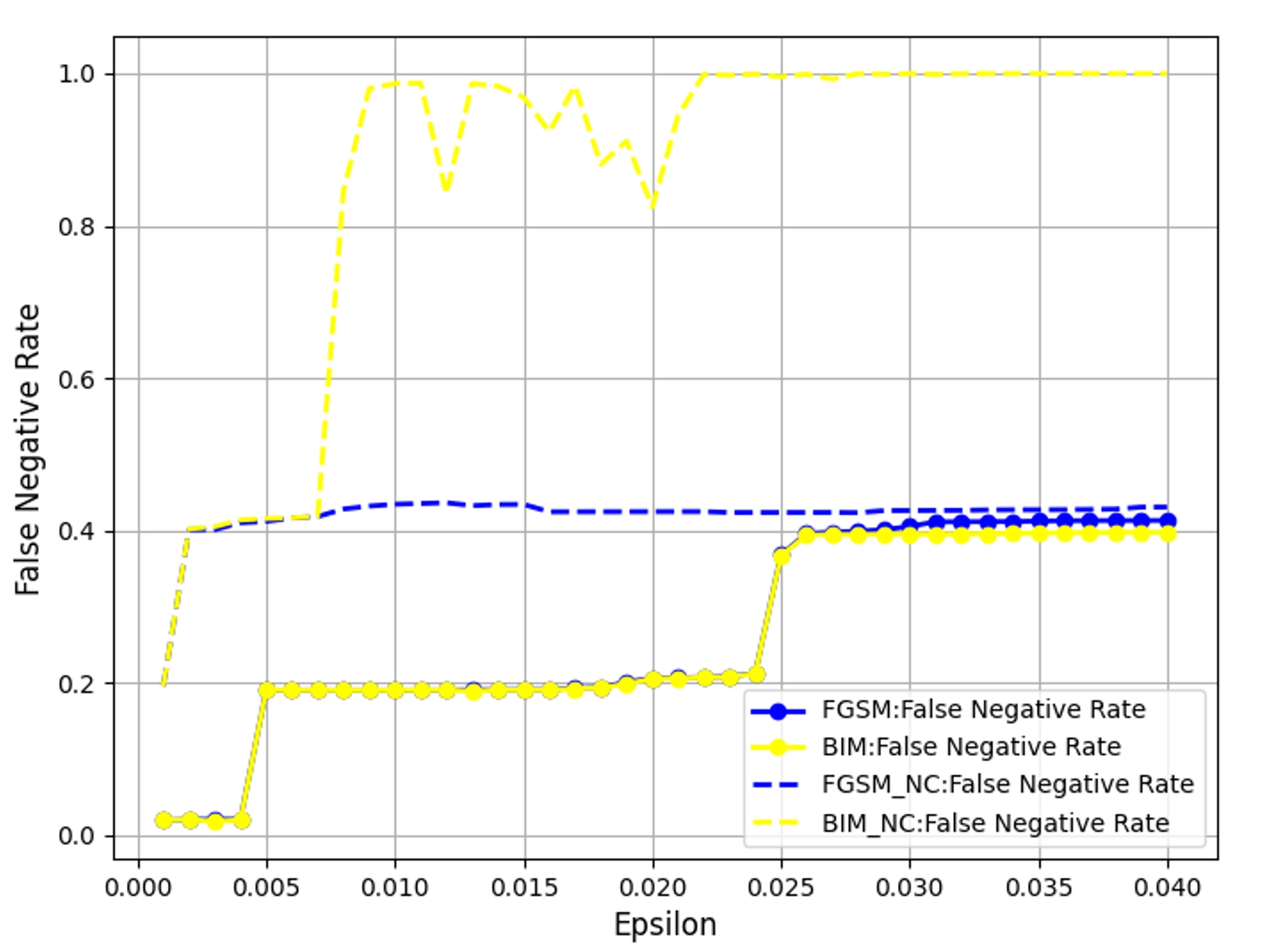}
        \label{fig:FGSM and BIM performance}
    } \hfill
    \subfigure[False Negative Rate (FNR) for proposed UAP]{
        \includegraphics[width=0.31\textwidth, height=0.18\textheight, keepaspectratio=false]{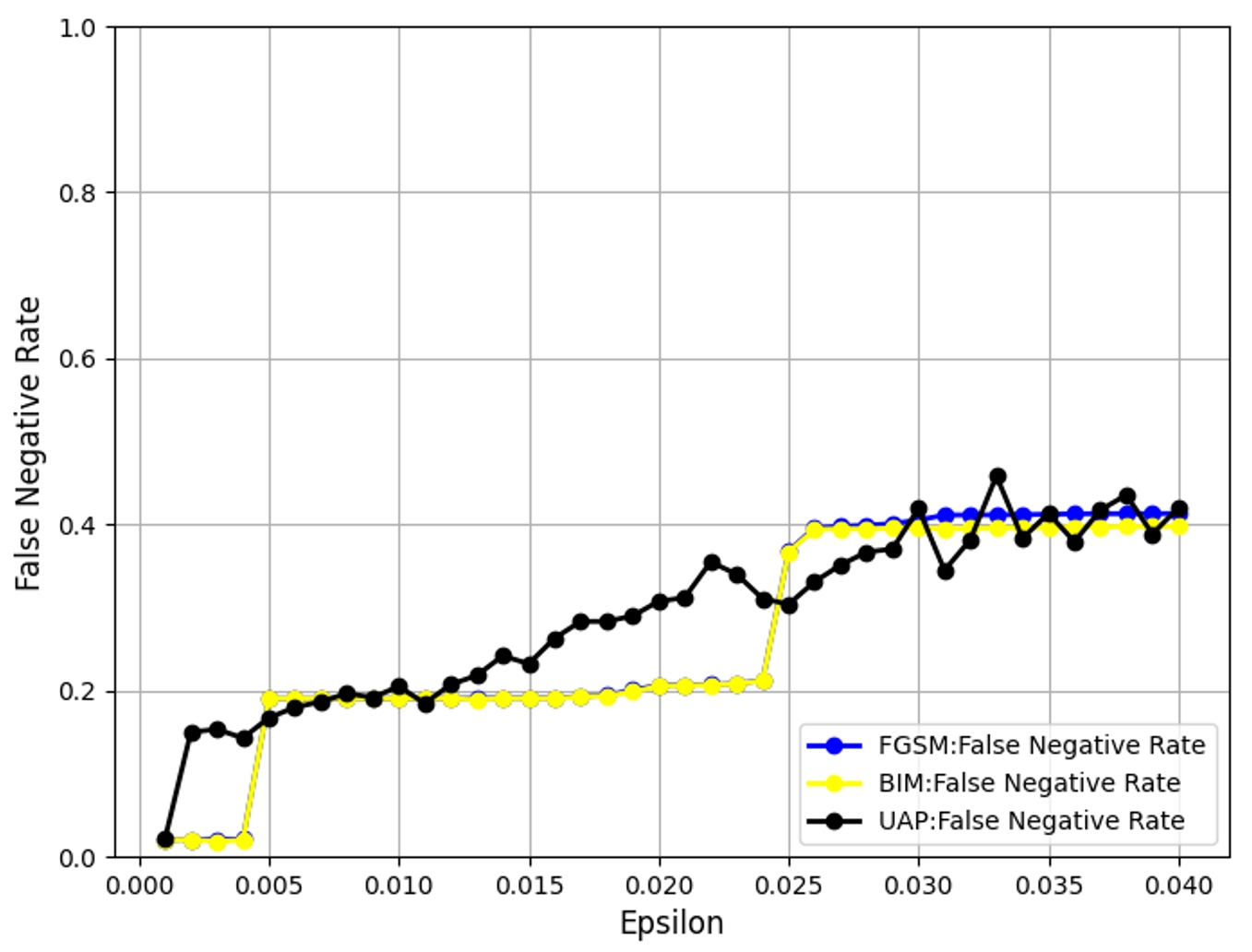}
        \label{fig:UAP FNR}
    } \hfill
    \subfigure[Accuracy for proposed UAP]{
        \includegraphics[width=0.31\textwidth, height=0.18\textheight, keepaspectratio=false]{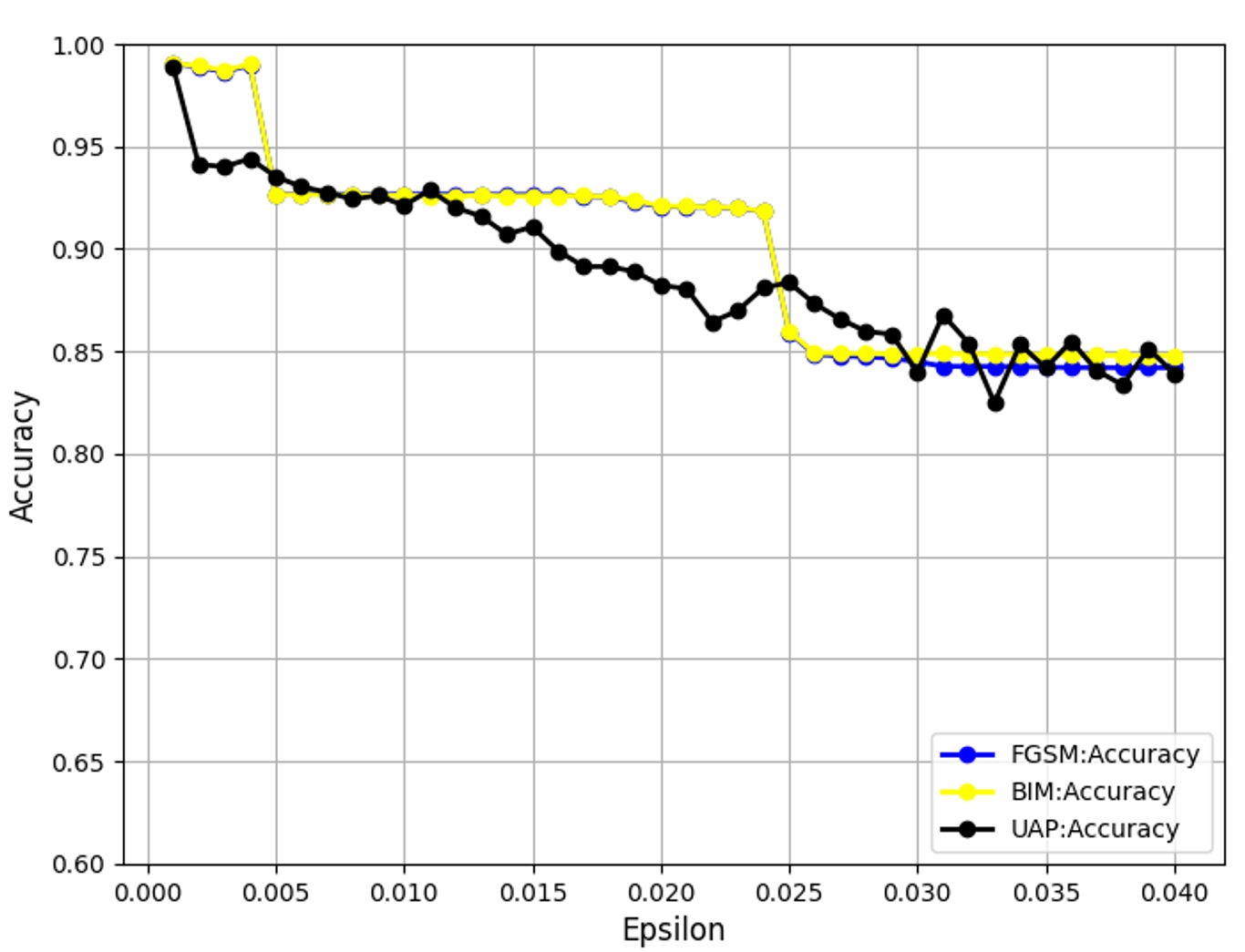}
        \label{fig:UAP Acc}
    }
    \caption{Attack performance of FGSM/BIM and proposed UAP}
    \label{fig:CombFGSM and UAP}
\end{figure*}
Each training session consists of 10 episodes and adopts an epsilon-greedy exploration strategy, where the exploration rate decays from 1.0 to 0.05 over the initial 10\% of training steps. The discount factor $\gamma$ is set to 0.001 as mentioned earlier. The network is optimized using the Adam optimizer with a learning rate of $1 \times 10^{-4}$. A total of 10 training runs were conducted, to ensure that the evaluation reflects typical rather than extreme performance. The agent with accuracy closest to the median among the 10 runs was chosen as the target model, with 99.78\% on the testing set, shown in Figure \ref{fig:DRL performance}. This result substantiates the effectiveness of both our data preprocessing strategy and the proposed DRL-based intrusion detection architecture.

All further adversarial attacks were implemented under a global setting constrained by the $L_\infty$-norm, with the perturbation $\epsilon$ ranging from 0 to 0.04 to ensure the \textit{imperceptible} for adversarial examples under the normalized feature space. For the two input-dependent attacks FGSM and BIM, apart from the description provided in Section \ref{section: constraint FGSM/BIM UAP}, BIM was further parameterized with 20 epsilon steps, resulting in a per-step size equal to $\epsilon / 20$, the maximum iterations number for BIM was capped at 100. Additionally, since the generation of UAP involves random seed set selection and shuffling, we report the average results over 80 independent runs for performance stability.
Figure \ref{fig:FGSM and BIM performance} illustrates the performance of FGSM and BIM attacks across the predefined epsilon value range. To highlight the impact of our proposed domain constraint, we additionally include the results of both attack methods without masking and recalculation (shown as dashed lines). It is evident that the domain constraint significantly affects the effectiveness of both attacks, causing a performance loss of over 50\% for BIM. The stricter constraints limit the modifiable feature space, which further emphasizes the necessity of adopting realistic settings in adversarial attack research for IDS.
\begin{figure*}[t]
\setlength{\abovecaptionskip}{1pt}
    \centering
    \subfigure[False Negative Rate (FNR)]{
        \includegraphics[width=0.31\textwidth, height=0.18\textheight, keepaspectratio=false]{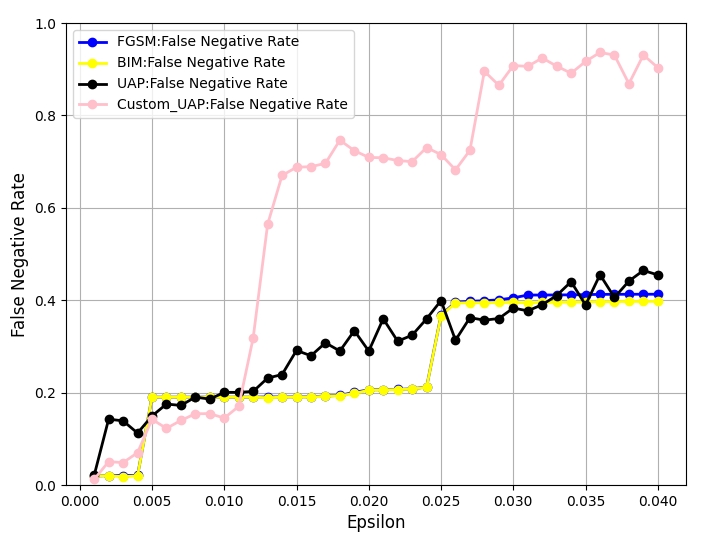}
        \label{fig:Custom_UAP performance}
    } \hfill
    \subfigure[Accuracy]{
        \includegraphics[width=0.31\textwidth, height=0.18\textheight, keepaspectratio=false]{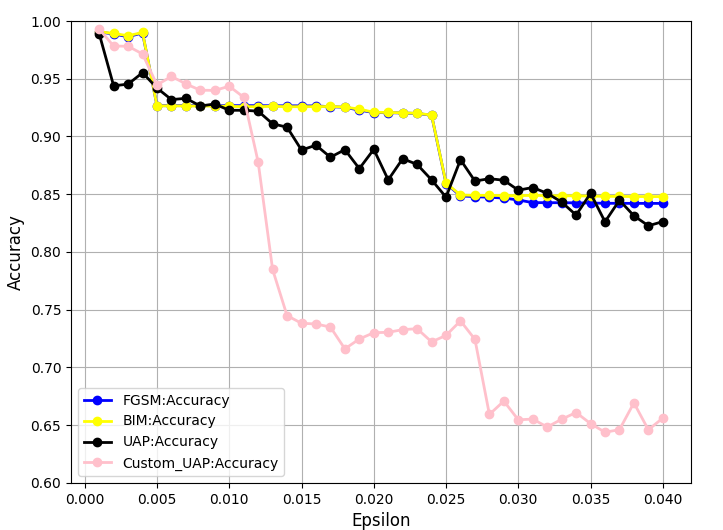}
        \label{fig:Custom_UAP performance Acc}
    } \hfill
    \subfigure[$\text{PCC}_{\text{pertu}}$]{
        \includegraphics[width=0.31\textwidth, height=0.18\textheight, keepaspectratio=false]{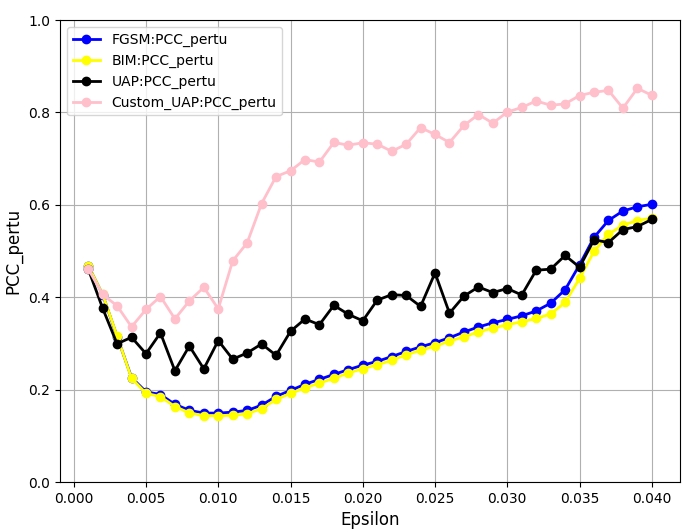}
        \label{fig:Custom_UAP performance PCC}
    }
    \caption{Performance of Customized UAP with previous three attacks.}
    \label{fig: Custom_UAP}
\end{figure*}

To evaluate the generalization capability of the proposed UAP attack, we conducted experiments using the UAP method described in \mbox{Algorithm \ref{alg:uap}} and benchmarked its performance against FGSM and BIM, with FNR shown in Figure \ref{fig:UAP FNR} and accuracy in \ref{fig:UAP Acc}, where the x-axis represents the selected $\epsilon$ values and the y-axis indicates the corresponding evaluation metrics. It can be found that the proposed UAP overall achieved comparable attack performance to the two previously introduced input-dependent methods in our case. The epsilon range can be partitioned into three regions, lower, middle, and higher, each showing distinct patterns. The proposed UAP attack achieves a better performance at a lower perturbation magnitude ($\epsilon \approx$ 0.003) with 15\% FNR  and around 93\% accuracy, whereas FGSM and BIM only show similar effects at $\epsilon \approx$ 0.005. In the mid-range ($\epsilon$ = 0.010–0.025), the proposed UAP outperformed both FGSM and BIM, with FNR rising steadily from 20\% to 35\% and accuracy decreasing from 93\% to 86\%, whereas FGSM and BIM maintained the same performance as at $\epsilon$ = 0.005 (FNR 20\%, accuracy 93\%). In the higher range ($\epsilon \approx$ 0.033–0.040), the proposed UAP achieved comparable performance to the two data-dependent methods (FNR 40\%, accuracy 85\%). Overall, the results indicate that the proposed UAP meets the adversarial objective, namely, manipulating the outcomes of malicious instances while benign instances stay unchanged, showing the threat of UAP attack in IDS domain. Unlike input-specific attacks such as FGSM and BIM, which require per-instance gradient computation, the generated UAP can be applied directly to diverse inputs without further optimization, greatly reducing computational cost and enabling realistic large-scale or time-sensitive attacks.

\subsection{Customized UAP Result}

Now we present the results for the Customized UAP and compare its performance against the three previously introduced attack methods. The FNR and accuracy results can be found in Figure \ref{fig:Custom_UAP performance} and Figure \ref{fig:Custom_UAP performance Acc}, respectively. A remarkable observation is that the proposed Customized UAP (in pink line) substantially outperforms the proposed UAP in most of the evaluated epsilon range. All four attack methods maintain a comparable level of effectiveness before $\epsilon = 0.010$, with FNR of 16\%-20\% and accuracy of 93\%-95\%, indicating limited divergence in their impact within the low perturbation range. From this point onward, Customized UAP shows a rapid increase in attack performance, achieving a FNR nearly twice (around 70\%) as high as the other three attacks, resulting in a significantly large number of instances classified as benign and the degradation of classification performance. After $\epsilon = 0.025$, it experiences a second surge in performance with the other three and reaches a peak rate of approximately 90\% for FNR and 65\% for accuracy at $\epsilon$ = 0.030, and remains stable until $\epsilon$ = 0.040. Contrastively, the other attack methods plateau around 40\% FNR under the same perturbation magnitude. Notably, within the mid-range interval ($\epsilon = 0.010$ to $0.025$), the newly proposed PCC-based loss function substantially enhances the UAP's performance (FNR from 20\% to 70\%, accuracy from 93\% to 73\%), demonstrating that it not only preserves the computational efficiency but also boosts the attack effectiveness of UAP.


The results of $\text{PCC}_{\text{pertu}}$ value are shown in Figure~\ref{fig:Custom_UAP performance PCC}. It further provides insight into the question we raised in Section \ref{sec:custom_uap}. Since the perturbation generated by \mbox{Customized UAP} is explicitly optimized to increase the $\text{PCC}_{\text{pertu}}$ value, it tends to dominate the final prediction outcome more than the perturbations produced by the other three attack methods, leading to better performance. This trend remains consistently observable within the selected $\epsilon$ range, which encompasses an early-stage decline region ($\epsilon=[0, 0.010]$) and a later-stage performance rising ($\epsilon=[0.010, 0.040]$). It is worth noting that within the epsilon range of 0.010 to 0.015, the $\text{PCC}_{\text{pertu}}$ value of the Customized UAP increases rapidly to approximately 0.7, showing high similarity between the predicted outcome of the perturbation and Customized UAP attack. Beyond this interval, it exhibits a steadily increasing trend and reaches 0.82 when $\epsilon = 0.040$, significantly outperforming the other three attack methods, whose $\text{PCC}_{\text{pertu}}$ are around 0.6. This further shows that our proposed Customized UAP, with using $\text{PCC}_{\text{pertu}}$ as the loss function enhances the attack performance in UAP generation compared to the conventional CrossEntropy objective of the proposed UAP attack.
\vspace{-5pt}
\begin{figure*}[t]
\setlength{\abovecaptionskip}{1pt}
    \centering
    \subfigure[False Negative Rate (FNR)]{
        \includegraphics[width=0.31\textwidth, height=0.18\textheight, keepaspectratio=false]{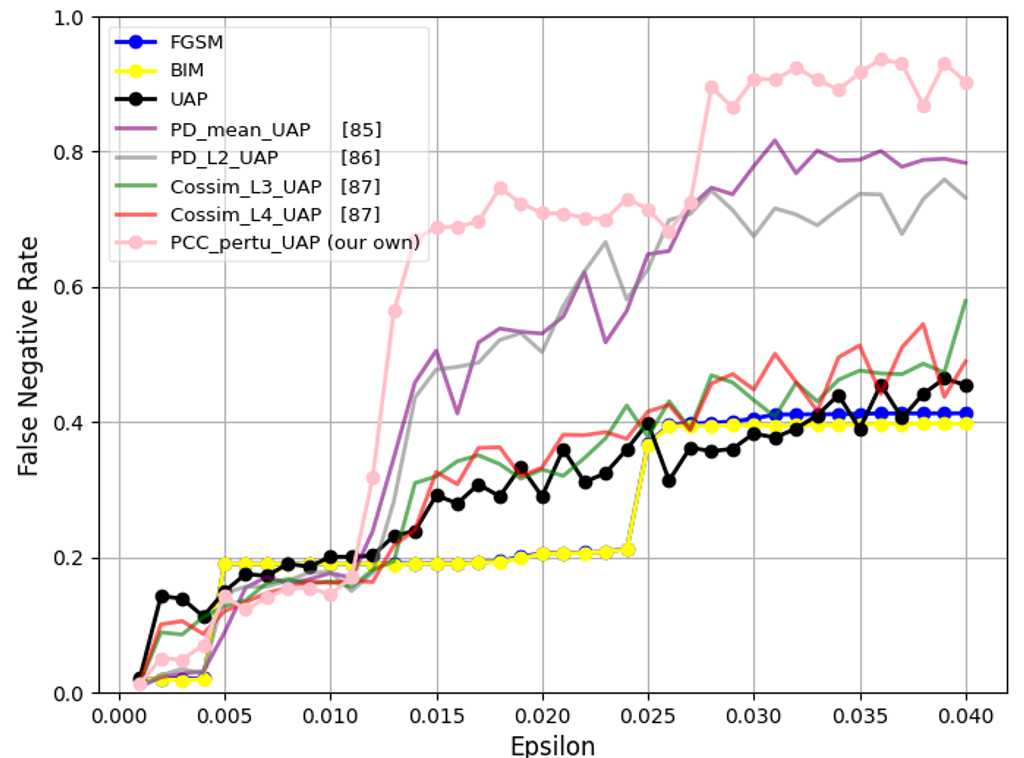}
        \label{fig:all cus_FNR}
    } \hfill
    \subfigure[Accuracy]{
        \includegraphics[width=0.31\textwidth, height=0.18\textheight, keepaspectratio=false]{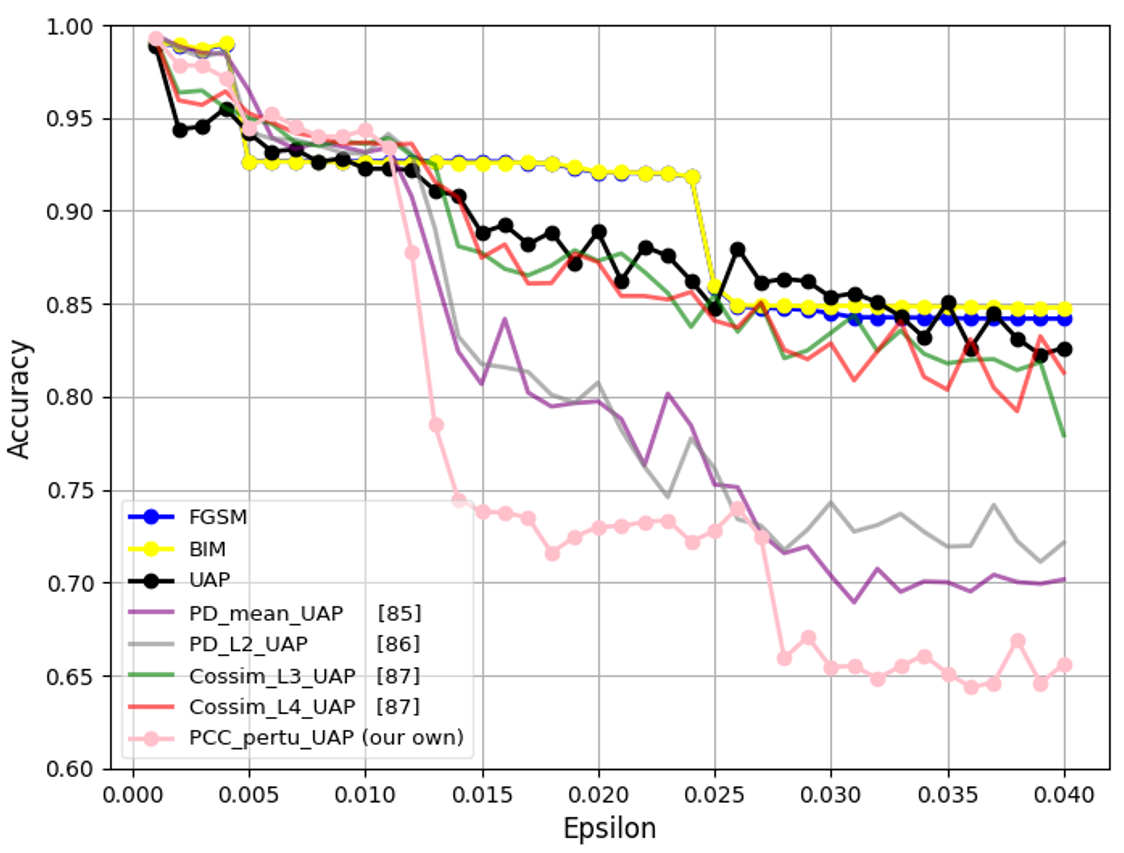}
        \label{fig:all cus_Acc}
    } \hfill
    \subfigure[$\text{PCC}_{\text{pertu}}$]{
        \includegraphics[width=0.31\textwidth, height=0.18\textheight, keepaspectratio=false]{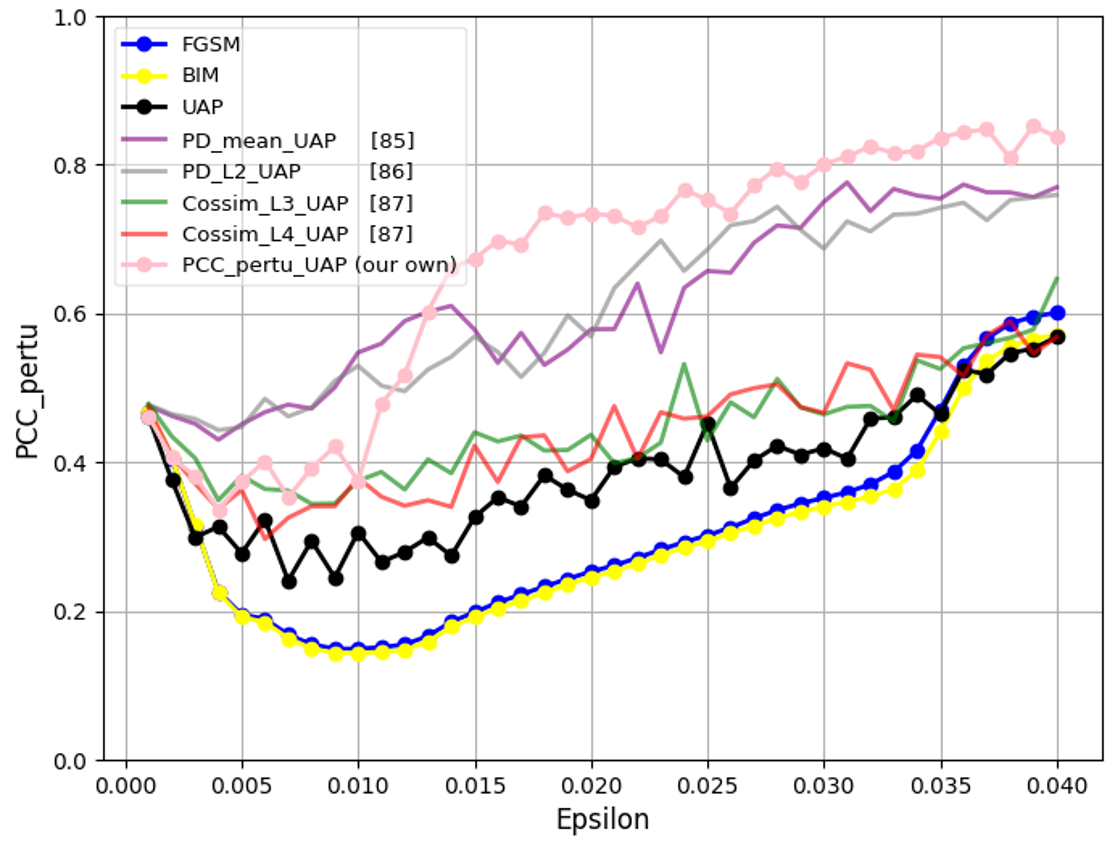}
        \label{fig:all cus_PCCx}
    }
    \caption{Performance comparison of custom candidate losses with three baseline method.}
    \label{fig:all cus}
\end{figure*}
\subsection{Comparison with other UAP Baselines}
The experimental results presented above demonstrate that the proposed Customized UAP outperforms both the original proposed UAP and the two input-dependent attacks for overall effectiveness. In order to provide a comprehensive analysis of attack performance, now we further introduce the result of other four established UAP baselines, whose details and index are shown in Table \ref{tab: Milti Loss} previously. Multiple performance metrics of all the attacks are shown in Figure \ref{fig:all cus}, including FNR, Accuracy and $\text{PCC}_{\text{pertu}}$ in \ref{fig:all cus_FNR}, \ref{fig:all cus_Acc} and \ref{fig:all cus_PCCx}. Beginning at the lowest $\epsilon$ value, all the established UAP baselines as well as our Customized UAP shows temporarily outperforming the proposed UAP attack. Their performance reaches the same level as the proposed UAP (black line) and the other two data-dependent attacks at $\epsilon = 0.010$, all yielding an FNR of 20\% and an accuracy of 93\%. 

Based on the performance after this point, the six candidate loss functions can be grouped into three tiers. Loss functions from work~\cite{ye2023fg} (denoted as \textit{COSSIM\_L3\_UAP}, \textit{COSSIM\_L4\_UAP} in Figure \ref{fig:all cus}) focus on reducing the similarity of activation result between the clean data and corresponding adversarial examples from different perspectives. However, such approaches do not lead to performance improvements for UAP in our case, with following the same trend as the black line in the selected epsilon range, keep holding the increasing of their attack performance and remain a stable level in $\epsilon=[0.025, 0.040]$, as shown across all three subfigures (about 45\% FNR, 82\% Accuracy and 0.6 $\text{PCC}_{\text{pertu}}$). Secondly, loss functions from study~\cite{mopuri2017fast} and~\cite{mopuri2018generalizable} (\textit{PD\_mean\_UAP} and \textit{PD\_L2\_UAP}) can be categorized as the middle-tier group. These losses aim to increase the overall activation of neurons in the hidden layers, thereby making it more difficult for the network to distinguish meaningful features and further causing the classifier to fail. This class of loss functions successfully improved the performance of UAPs in the mid-to-late epsilon range, from $\epsilon$ = 0.010 to 0.030, and keep the performance level at around 75\% FNR and 72\% accuracy until the upper bound $\epsilon$ = 0.040. It can be seen that the tier 2 UAP baselines successfully outperform the first group, leading to more misclassification of positive samples while maintaining correct classification of negative samples. 

Among all the candidate loss functions, our proposed Customized UAP (\textit{PCC\_pertu\_UAP}) clearly outperforms all other attack methods within the entire epsilon range, positioning itself as the top-tier approach, and finally achieves the performance of around 90\% FNR and 65\% accuracy. As shown in Figure \ref{fig:all cus_PCCx}, the perturbations it generates are also the most dominant in influencing model predictions of the corresponding adversarial examples, with around 0.82 of $\text{PCC}_{\text{pertu}}$ at $\epsilon=0.040$.  After analyzing the experiment results, we can conclude that the proposed Customized UAP outperforms the four UAP baselines, which improves both the computational efficiency and the attack effectiveness of adversarial attacks on IDS.
\section{Conclusion} \label{Chap:5}
In this work, we propose a novel and practical UAP attack against a DRL-based IDS under realistic domain-specific constraints. To the best of our knowledge, this is the first work that focuses on developing a UAP against a DRL-based IDS under realistic domain constraints based on not only the basic rules of the domain but also mathematical relations between the features. In addition, we further enhance the attack performance by integrating the PCC value into UAP design. This is also the first work using the PCC value in the UAP generation process, even in the broader context. To comprehensively evaluate the performance of the proposed attacks, we compare the attack performance with four established UAP baselines as well as two input-dependent adversarial attacks: FGSM and BIM. Our experimental results demonstrate the proposed UAP achieves a comparable effectiveness to two data-dependent methods, validating its viability under practical limitations. Furthermore, our Customized UAP attack outperforms not only our proposed UAP attack, but also four other established UAP baselines.
\vspace{-2pt}
\section*{Acknowledgments}
This work is supported by UKRI through the research grants EP/R007195/1 (Academic Centre of Excellence in Cyber Security Research - University of Warwick) and EP/Y028813/1 (National Hub for Edge AI).

\bibliographystyle{IEEEtran}
\bibliography{refs}

\end{document}